\documentclass[aps,pre,twocolumn,showpacs,floatfix,preprintnumbers,superscriptaddress,amsmath,amssymb,groupedaddress]{revtex4-1}
\usepackage{amsmath}
\usepackage{amssymb}
\usepackage{graphicx}
\usepackage{epsfig}
\usepackage{dcolumn}
\usepackage{bm}
\usepackage{array}
\usepackage{xcolor}
\usepackage{subfig}
\usepackage{float}
\usepackage{physics}
\usepackage{graphicx}
\usepackage{dcolumn}
\usepackage{bm}

\usepackage{dsfont,bbold}

\begin{document}

\title{Distribution of the ratio of two consecutive level spacings in orthogonal to unitary crossover ensembles}

\author{Ayana Sarkar}
\email{ayana1994s@gmail.com}
\author{Manuja Kothiyal}
\author{Santosh Kumar}
\email{skumar.physics@gmail.com}
\affiliation{Department of Physics, Shiv Nadar University, Gautam Buddha Nagar, Uttar Pradesh 201314, India}

\begin{abstract}
The ratio of two consecutive level spacings has emerged as a very useful metric in investigating universal features exhibited by complex spectra. It does not require the knowledge of density of states and is therefore quite convenient to compute in analyzing the spectrum of a general system. The Wigner-surmise-like results for the ratio distribution are known for the invariant classes of Gaussian random matrices. However, for the crossover ensembles, which are useful in modeling systems with partially broken symmetries, corresponding results have remained unavailable so far. In this work, we derive exact results for the distribution and average of the ratio of two consecutive level spacings in the Gaussian orthogonal to unitary crossover ensemble using a $3\times 3$ random matrix model. This crossover is useful in modeling time-reversal symmetry breaking in quantum chaotic systems. Although based on a $3\times 3$ matrix model, our results can also be applied in the study of large spectra, provided the symmetry-breaking parameter facilitating the crossover is suitably scaled. We substantiate this claim by considering Gaussian and Laguerre crossover ensembles comprising large matrices. Moreover, we apply our result to investigate the violation of time-reversal invariance in the quantum kicked rotor system. 
\end{abstract}

\maketitle

\section{Introduction}

Since Wigner's pioneering work in the 1950s, random matrix theory (RMT) has developed into an important field dedicated to the statistical investigation of complex systems. From nuclear and particle physics~\cite{BFFMPW1981,MRW2010,SV1993,Verbaarschot1994,GW1997,Verbaarschot2011,Akemann2017} to finance~\cite{LCBP1999, PGRAS1999,MS1999}, from classical and quantum information theory~\cite{TV2004,CD2011,KP2010c,ZS2001,SZ2004,BL2002,Majumdar2011,KP2011a,KSA2017} to transport phenomena in mesoscopic systems~\cite{Beenakker1997,Beenakker2015}, RMT has found widespread application in a broad range of areas~\cite{Mehta2004,GMW1998,Handbook2011,Forrester2010,Haake2010}. One of the fascinating aspects of RMT is its prediction of certain universal features which are found to be shared by a wide variety of completely unrelated systems. It was conjectured in a seminal work by Bohigas {\it et al.} that the local fluctuation properties of the spectra of quantum chaotic systems coincide with those of random matrices~\cite{BGS1984} and there has been overwhelming evidence in favor of this conjecture~\cite{BGS1984,Mehta2004,GMW1998,Handbook2011,Forrester2010,Haake2010}. In contrast, as was asserted by Berry and Tabor, integrable systems conform to Poissonian spectral fluctuations~\cite{BT1977}.

Dyson's three fold classification scheme involves three invariant classes of random matrices, which apply depending on the time-reversal and rotational symmetries exhibited by a system~\cite{Dyson1962,Mehta2004,GMW1998,Forrester2010,Haake2010,Handbook2011}. In the classical Gaussian case, these are Gaussian orthogonal ensemble (GOE), Gaussian unitary ensemble (GUE), and Gaussian symplectic ensemble (GSE). Additionally, there are instances when a symmetry is \emph{partially broken} and correspondingly the spectra exhibit statistics intermediate between two invariant classes. Such systems are modeled by crossover random matrix ensembles which interpolate between two symmetry classes as a certain symmetry-breaking or crossover parameter is varied~\cite{Dyson1972,PM1983,MP1983,PS1991,Pandey1981,Guhr1996,LH1991,LZS1991,FNH1999,NF2003,KP2009,KP2011,CK2013,CDK2014,Kumar2019,Mehta2004,Forrester2010,SMW2017}. The intermediate cases are realized when the crossover parameter assumes values between its two extremes. This crossover parameter can often be related to some physical quantity when modeling a quantum chaotic system. For instance, application of a weak magnetic field on a mesoscopic system (e.g., a quantum dot) leads to a partial time-reversal invariance violation~\cite{Beenakker1997}. Correspondingly, along with the change in spectral behavior, one also observes the effect on electronic transport properties leading to phenomena such as magnetoconductance~\cite{Beenakker1997,FP1995,BHMH2006,BC2007,KP2010a,KP2010b}. In this case, the crossover parameter governing the orthogonal to unitary transition can be related to the magnetic flux. Another prominent example of such crossovers is the one involving Poisson to RMT spectral fluctuations. It is used to model transition from integrability to chaos and has been used in several contexts~\cite{Brody1973,BR1984,Izrailev1990,MNS1994,RDP1999,PH2010,CFIM2012,CDK2014,LLA2015,BG2016,KRBL2010,CARK2012,IORH2013,RCCKLR2017,CK2013}. Interestingly, crossover ensembles also find applications in problems where the above-mentioned symmetries do not have any direct meaning, for example, in multiple channel wireless communication where the crossover is governed by a certain signal-fading parameter~\cite{KP2010c}. 

Nearest-neighbor spacing distribution (NNSD) is an important and widely studied statistical measure to quantify the local fluctuation properties of a given spectrum. As a matter of fact, the agreement of NNSD with the RMT result or the Poissonian statistics is considered as a litmus test to decide whether a system under consideration is chaotic or integrable. The NNSD expressions for the classical Gaussian ensembles based on $2\times 2$ matrices, famously known as \emph{Wigner surmise}, approximate the corresponding exact results for large matrices very closely and therefore are of immense usefulness in studying spectra of complex systems. Similarly, for the crossover ensembles, several Wigner-surmise-like results exist for the spacing distribution and apply to large spectra once the crossover parameter is suitably scaled~\cite{Brody1973,BR1984,HMF1988,Izrailev1989,CIM1991,AM1996,KS1999,SBW2012,SLMW2017,SMW2017}. Unfortunately, the empirical calculation of NNSD for a general system involves the difficult task of unfolding the spectrum which requires knowledge of the density of states to a very good accuracy, which is not always possible to acquire. Consequently, one seeks alternatives to NNSD which can be used to quantify the local fluctuation behavior without the need for unfolding. 

The ratio $r_n=s_n/s_{n-1}$ of two consecutive level spacings $s_n$ and $s_{n-1}$ is a quantity which is independent of the local density of states and has become quite popular in recent times. Here, the spacing is calculated as $s_{n}= x_{n+1}-x_{n}$, with $\{x_1,x_2,...\}$ being the ordered set of levels. The ratio of adjacent spacings was introduced by Oganesyan and Huse to study the spectral statistics of a one-dimensional tight-binding model of strongly interacting spinless fermions in a random potential~\cite{OH2007}. It was shown that spectral statistics of finite-sized samples show a crossover from GOE statistics in the diffusive regime at weak random potential to Poisson statistics in the localized regime at strong randomness. For their analysis, Oganesyan and Huse considered not $r_n$ but the related quantity $\tilde{r}_{n}=\min(s_{n},s_{n-1})/\max(s_{n},s_{n-1}) = \min(r_{n} ,1/r_{n})$. In Refs.~\cite{ABGR2013, ABGVV2013}, Atas {\it et al.} derived the probability density function (PDF) of the adjacent spacing ratio for the classical Gaussian ensembles and additionally for the Poissonian case which applies to spectra of integrable systems. Although these results are based on $3\times 3$ random matrices, similar to the Wigner surmise, they have been found to describe the fluctuation behavior of large spectra to a reasonable accuracy.

Due to the above-indicated advantage of the ratio distribution over the spacing distribution, it has gradually found widespread application in analyzing spectra of complex systems. This includes systems ranging from spin chains~\cite{OPH2009,PH2010,CFIM2012,CDK2014,LLA2015,BG2016}, Bose-Hubbard model~\cite{KRBL2010,CARK2012}, many-body quasiperiodic system~\cite{IORH2013}, molecular resonances~\cite{RCCKLR2017}, and fermions or bosons described by embedded random matrix ensembles~\cite{CK2013}. One of the key aspects considered in these works has been the Poisson-GOE crossover, for which a detailed analysis has been done by Kota and co-workers for the ratio distribution in Refs.~\cite{CK2013,CDK2014}. An empirical formula for ratio distribution in this crossover has also been provided in Ref.~\cite{CK2013}. The ratio distribution has also been used to examine the quantum chaos transition in a two-site Sachdev-Ye-Kitaev model, which is dual to an eternal traversable wormhole~\cite{GNRV2019}. Furthermore, exact distribution for the spacing ratio has been derived in Ref.~\cite{TKS2018} for distinguishing between localized and random states in quantum chaotic systems. Higher-order spacing ratios have been also explored in ~Refs. \cite{ABGVV2013,TBS2018,BTS2018} and utilized for studying the spectral fluctuations in complex quantum systems as well as those observed in empirical correlation matrices constructed using stock market and atmospheric pressure data. In addition, very recently, a phenomenological formula for the ratio distribution has been proposed in Ref.~\cite{CR2019} for several symmetry crossovers.

In this paper, we focus on an ensemble interpolating between GOE and GUE, and derive exact distribution for the ratio of consecutive level spacings $r$ as well as for $\tilde{r}=\min(r,1/r)$ using a $3\times 3$ random matrix model. Moreover, we give exact results for the corresponding averages $\langle r \rangle$ and $\langle \tilde{r} \rangle$. Similar to the Wigner surmise, our result can be applied to large dimension cases and to non-Gaussian random matrix ensembles once the crossover parameter is properly scaled. We substantiate this claim by considering interpolating ensembles of Gaussian and Wishart-Laguerre ensembles comprising large matrices. In quantifying the level fluctuation behavior of an arbitrary quantum chaotic system, our result can be applied by relating the crossover parameter of the RMT model to the time-reversal symmetry-breaking parameter of the given system. We demonstrate this by investigating the time-reversal symmetry breaking in a quantum kicked rotor (QKR) system~\cite{CCFI1979}. 

The rest of this paper is organized as follows. In Sec.~\ref{MMG}, we present the random matrix model used to capture the GOE-GUE crossover. In Sec.~\ref{rDist}, we derive exact results for the PDFs of $r$ and $\tilde{r}$, and the corresponding averages. In Sec.~\ref{scaling}, we examine the scaling behavior of the crossover parameter for Gaussian ensembles in large dimension cases. Section~\ref{ALE} deals with the investigation of orthogonal to unitary crossover in the Wishart-Laguerre ensemble of random matrices. In Sec.~\ref{QKR}, we apply our analytical result for studying the time-reversal symmetry breaking in a quantum kicked rotor system. We conclude with a brief summary in Sec.~\ref{discsum}.



\section{Matrix model for the GOE-GUE crossover}
\label{MMG}

We use the random matrix model proposed by Pandey and Mehta for GOE-GUE crossover~\cite{PM1983, MP1983}:
\begin{equation}
\label{PMmatrix1}
H=\sqrt{1-\alpha^2}\,H_1+\alpha H_2.
\end{equation}
In this equation, $H_1$ and $H_2$ are real-symmetric and complex-Hermitian random matrices with probability densities
\begin{equation}
\mathcal{P}(H_\beta)\propto \exp\left(-\frac{1}{4v^2}\tr H_\beta^2\right),~~~\beta=1,2,
\end{equation}
respectively. The parameter $v$ appearing above fixes the scale of the problem. Clearly, $H_1$ and $H_2$ belong to the GOE and GUE, respectively. The variances of the diagonal and off-diagonal parts (both real and imaginary for GUE) of the Gaussian matrix elements in either case are $2v^2$ and $v^2$, respectively. The matrix model in Eq.~\eqref{PMmatrix1} interpolates between GOE and GUE as the crossover parameter $\alpha$ is varied from 0 to 1. Other equivalent forms for the above matrix model are also possible. For example, Lenz and co-workers have considered the random matrix model~\cite{LH1991,LZS1991}
\begin{equation}
\label{PMmatrix2}
H=\frac{1}{\sqrt{1+\lambda^2}} H_1+\frac{\lambda}{\sqrt{1+\lambda^2}} H_2.
\end{equation}
The relation between parameters $\alpha$ and $\lambda$ can be readily established to be $\alpha=\lambda/\sqrt{1+\lambda^2}$ or $\lambda=\alpha/\sqrt{1-\alpha^2}$. The parameter $\lambda$ gives GOE and GUE in the limits $0$ and $\infty$, respectively. The joint eigenvalue density for the matrix model in Eq.~\eqref{PMmatrix1} is used in the next section to obtain an exact result for the distribution of ratio of consecutive level spacings.


\section{Exact ratio distribution for $3$-dimensional Gaussian matrices}
\label{rDist}
Considering the three-dimensional case, the joint probability density function for the orthogonal-unitary crossover in the Gaussian ensemble is given by~\cite{PM1983, MP1983}
\begin{align}
\label{jpdf}
\nonumber 
&P(x_1,x_2,x_3)= \frac{1}{48\sqrt{2}\,\pi v^6(1-\alpha^2)^{3/2}} \\ 
\nonumber
&\times  [f(x_1-x_2)-f(x_1-x_3)+f(x_2-x_3)]\\
&\times(x_1-x_2)(x_2-x_3)(x_1-x_3)\,e^{-\frac{1}{4v^2}(x_{1}^{2}+x_{2}^{2}+x_{3}^{2})},
\end{align}
where 
\begin{equation}
f(u)=\erf\left[\left(\frac{1-\alpha^2}{8\alpha^2 v^2}\right)^{1/2}u\right],
\end{equation} 
with $\erf(z)=(2/\sqrt{\pi})\int_0^z e^{-t^2}dt$ being the error function.
The limits $\alpha\to0$ and $\alpha\to 1$ in Eq.~\eqref{jpdf} lead to the joint probability densities of eigenvalues for GOE and GUE, respectively. We have
\begin{align}
\nonumber
P_\text{GOE}(x_1,x_2,x_3)=\frac{1}{48\sqrt{2}\pi v^6}\,e^{-\frac{1}{4v^2}(x_{1}^{2}+x_{2}^{2}+x_{3}^{2})}\\
\times|(x_1-x_2)(x_1-x_3)(x_2-x_3)|,
\end{align}
\begin{align}
\nonumber
P_\text{GUE}(x_1,x_2,x_3)=\frac{1}{768\pi^{3/2}v^9}\,e^{-\frac{1}{4v^2}(x_{1}^{2}+x_{2}^{2}+x_{3}^{2})}\\
\times[(x_1-x_2)(x_1-x_3)(x_2-x_3)]^2.
\end{align}

For the purpose of computing the desired ratio distribution, we order the eigenvalues such that $x_1\leq x_2\leq x_3$. The ratio of consecutive level spacings is then given by
\begin{equation}
r=\frac{x_3-x_2}{x_2-x_1}.
\end{equation}
The probability density function of $r$ can be obtained as
\begin{align}
\label{pr0}
\nonumber
p(r)=3! \int_{-\infty}^{\infty}dx_2 \int_{-\infty}^{x_2} dx_1\int_{x_2}^{\infty} dx_3 \,P(x_1,x_2,x_3)\\
\times \delta\left(r-\frac{x_3-x_2}{x_2-x_1}\right).
\end{align}
We proceed similarly to Refs.~\cite{ABGR2013,ABGVV2013} and implement the change of variables $(x_1,x_2,x_3)\rightarrow(x,x_2,y)$ with $x$ = $x_2-x_1$, $y$ = $x_3-x_2$. This gives us
\begin{align}
\nonumber
p(r)& =6 \int_{-\infty}^{\infty}dx_2\int_0^\infty dx\int_0^\infty dy \,P(x_2-x,x_2,x_2+y)\,\\ \nonumber&\times\delta\left(r-\frac{y}{x}\right).
\end{align}
On using Eq.~\eqref{jpdf}, we find that in the above expression, the variable $x_2$ appears only in the exponential factor and therefore the integral over it can be trivially performed, leading us to
\begin{align}
\label{integral}
 \nonumber
&p(r)=\frac{1}{4\sqrt{6\pi}\, v^5(1-\alpha^2)^{3/2}}\int_{0}^{\infty}dx \,\int_{0}^{\infty}dy\,x\, y\,(x+y)\\
&\times[f(x)+f(y)-f(x+y)]\,e^{-\frac{1}{6v^2}(x^2+xy+y^2)}\,\delta\Big(r-\frac{y}{x}\Big).
\end{align}
From the above equation, the joint density of consecutive spacings $x$ and $y$ can be read as
\begin{align}
\label{jpdsp}
\nonumber
&\widehat{P}(x,y)=\frac{1}{4\sqrt{6\pi}\, v^5(1-\alpha^2)^{3/2}}\,x\, y\,(x+y)\\
&\times[f(x)+f(y)-f(x+y)]\,e^{-\frac{1}{6v^2}(x^2+xy+y^2)},
\end{align}
which is symmetric in $x$ and $y$. Now, using the result $\delta(r-y/x)= x\,\delta(y-rx)$, we get rid of the $y$ integral from Eq.~\eqref{integral}  and are left with an integration on $x$ only,
\begin{align}
\nonumber 
&p(r)=\frac{r(r+1)}{4\sqrt{6\pi}\, v^5(1-\alpha^2)^{3/2}}\int_{0}^{\infty} dx\,x^4\\
&\times[f(x)+f(rx)-f(r+rx)]\,e^{-\frac{x^2}{6v^2} (r^2+r+1)}.
\end{align}
The above expression involves three integrals of identical structure which offers an exact result,
\begin{align}
\label{getazeta}
\nonumber
g(\eta,\zeta)& := \frac{4\sqrt{\pi}}{v^5}\int_{0}^{\infty}\, dx\,x^4\,\exp\left(-\frac{\eta^2 x^2}{v^2}\right)  \erf\left(\frac{\zeta x}{v}\right)\\
&= \frac{\zeta(5\eta^2+3\zeta^{2})}{\eta^4(\eta^2+\zeta^{2})^2}+\frac{3}{\eta^5}\arctan\!\left(\frac{\zeta}{\eta}\right).
\end{align}
Using the above integration result, we obtain the final expression for the probability density function for the ratio of two consecutive level spacings as
\begin{equation}
\label{pr}
p(r)= \frac{r(r+1)}{16\sqrt{6}\,\pi(1-\alpha^2)^{3/2}}
\left[g(a,b)+g(a,br)-g(a,br+b)\right],
\end{equation}
where $g(\eta,\zeta)$ is as given in Eq.~\eqref{getazeta} and 
\begin{eqnarray}
\label{ab}
a=\sqrt{\frac{r^2+r+1}{6}},~
b=\sqrt{\frac{1-\alpha^2}{8\alpha^2}}.
\end{eqnarray}
We note that the function $g(\eta,\zeta)$, and hence $p(r)$, is independent of $v$. This is expected, as ratio distribution has to be independent of the global scale of the spectrum. This can be seen from Eq.~\eqref{pr0} also by considering the scaling $x \rightarrow vx$. The three ``arctan" terms in $p(r)$ as in Eq.~\eqref{pr} may be combined to yield a single term, giving us, overall,
\begin{align}
\label{pr1}
\nonumber
p(r)&= \frac{r(r+1)}{16\sqrt{6}\,\pi(1-\alpha^2)^{3/2}}
\Bigg[\frac{b(5a^2+3b^2)}{a^4(a^2+b^2)^2}\\
\nonumber
&+\frac{br(5a^2+3b^2r^2)}{a^4(a^2+b^2r^2)^2}-\frac{b(r+1)(5a^2+3b^2(r+1)^2)}{a^4(a^2+b^2(r+1)^2)^2}\\
&+\frac{3}{a^5}\arctan\left(\frac{b^3r(r+1)}{a^3+ab^2(r^2+r+1)}\right)\Bigg].
\end{align}
Equations~\eqref{pr} and~\eqref{pr1} may be written in terms of the parameter $\lambda$ of the matrix model given in Eq.~\eqref{PMmatrix2} by replacing $1/(1-\alpha^2)^{3/2}$ by $(1+\lambda^2)^{3/2}$ and $b$ in Eq.~\eqref{ab} by $1/(\sqrt{8}\,\lambda)$. 
In the limit $\alpha\rightarrow 0$, we have $b\rightarrow\infty$, and therefore $g(a,b)=g(a,br)=g(a,br+b)=3\pi/(2a^5)$. Consequently, we obtain the result for GOE~\cite{ABGR2013,ABGVV2013}:
\begin{equation}
\label{pGOE}
p_\mathrm{GOE}(r)=\frac{27}{8}\frac{r(r+1)}{(r^2+r+1)^{5/2}}.
\end{equation}
Now, for $\zeta\rightarrow0$, we obtain $g(\eta,\zeta)=8\zeta/\eta^6-8\zeta^3/\eta^8+\mathcal{O}(\zeta^5)$ by invoking the expansion of $\arctan z$ about $z=0$. Hence, in the limit of $\alpha\rightarrow 1$ or $b\rightarrow 0$, we have 
\begin{align}
\nonumber
&g(a,b)+g(a,br)-g(a,br+b)
=\frac{8b}{a^6}-\frac{8b^3}{a^8}+\frac{8br}{a^6}\\
\nonumber
&-\frac{8b^3r^3}{a^8}-\frac{8(r+1)b}{a^6}+\frac{8(r+1)^3b^3}{a^8}+\mathcal{O}(b^5)\\
\nonumber
&=\frac{24r(r+1)b^3}{a^8}+\mathcal{O}(b^5)\\
&=\frac{24r(r+1)(1-\alpha^2)^{3/2}}{8^{3/2}a^8}+\mathcal{O}\big((1-\alpha^2)^{5/2}\big).
\end{align}
This leads to GUE result in the limit $\alpha\rightarrow1$~\cite{ABGR2013,ABGVV2013}:
\begin{equation}
\label{pGUE}
p_\mathrm{GUE}(r)=\frac{81\sqrt{3}}{4\pi}\frac{r^2(r+1)^2}{(r^2+r+1)^4}.
\end{equation}
\begin{figure}[h!]
\centering
\includegraphics[width=0.95\linewidth]{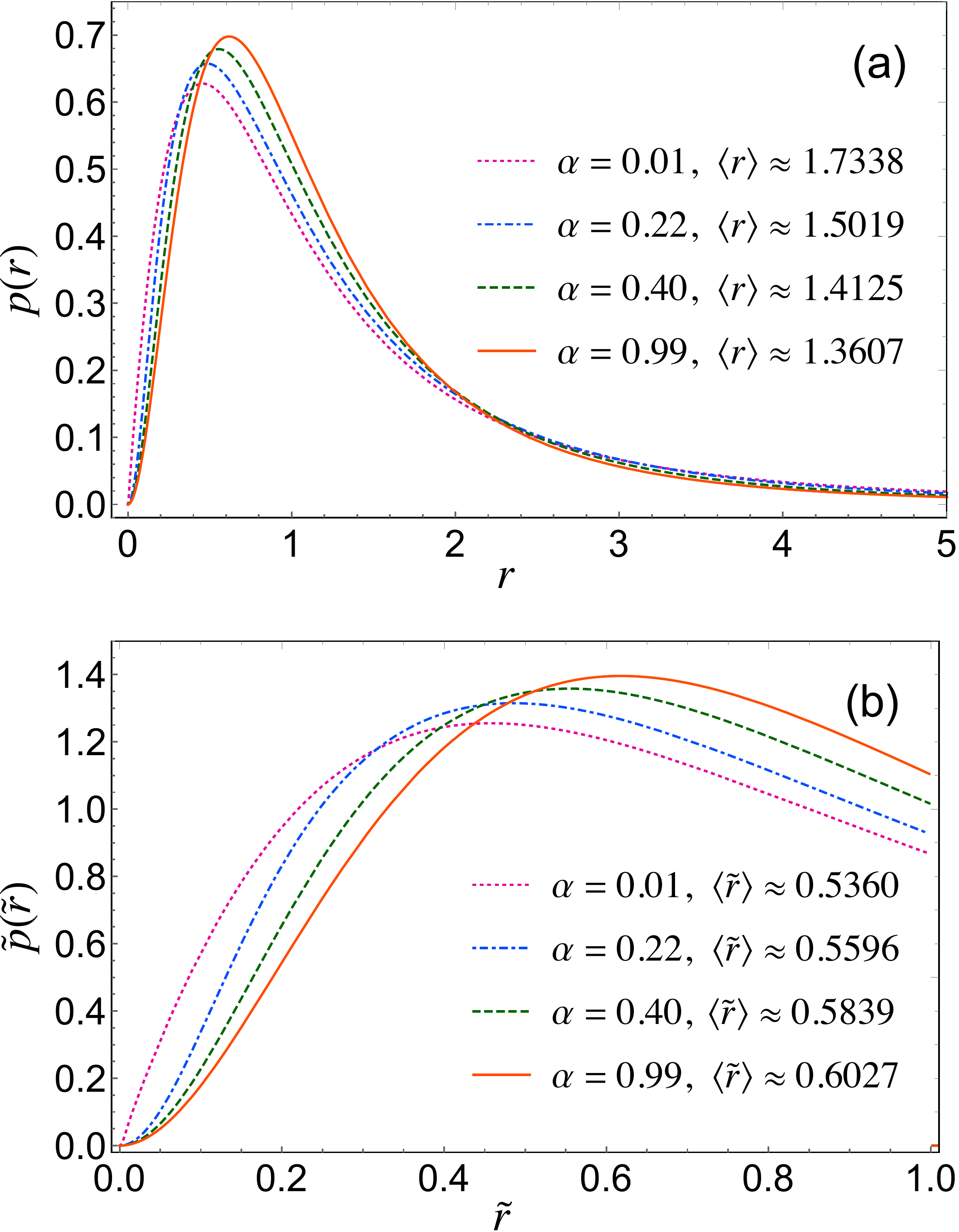}\\
\caption{Probability density functions (a) $p(r)$ and (b) $\tilde{p}(\tilde{r})$, as in Eqs.~\eqref{pr1}, and~\eqref{prtilde} for different $\alpha$ values. The corresponding average values $\langle r \rangle$ and $\langle \tilde{r} \rangle$ are also mentioned in each case.}
\label{fig1}
\end{figure}
\begin{figure*}[ht!]
\centering
\includegraphics[width=0.9\linewidth]{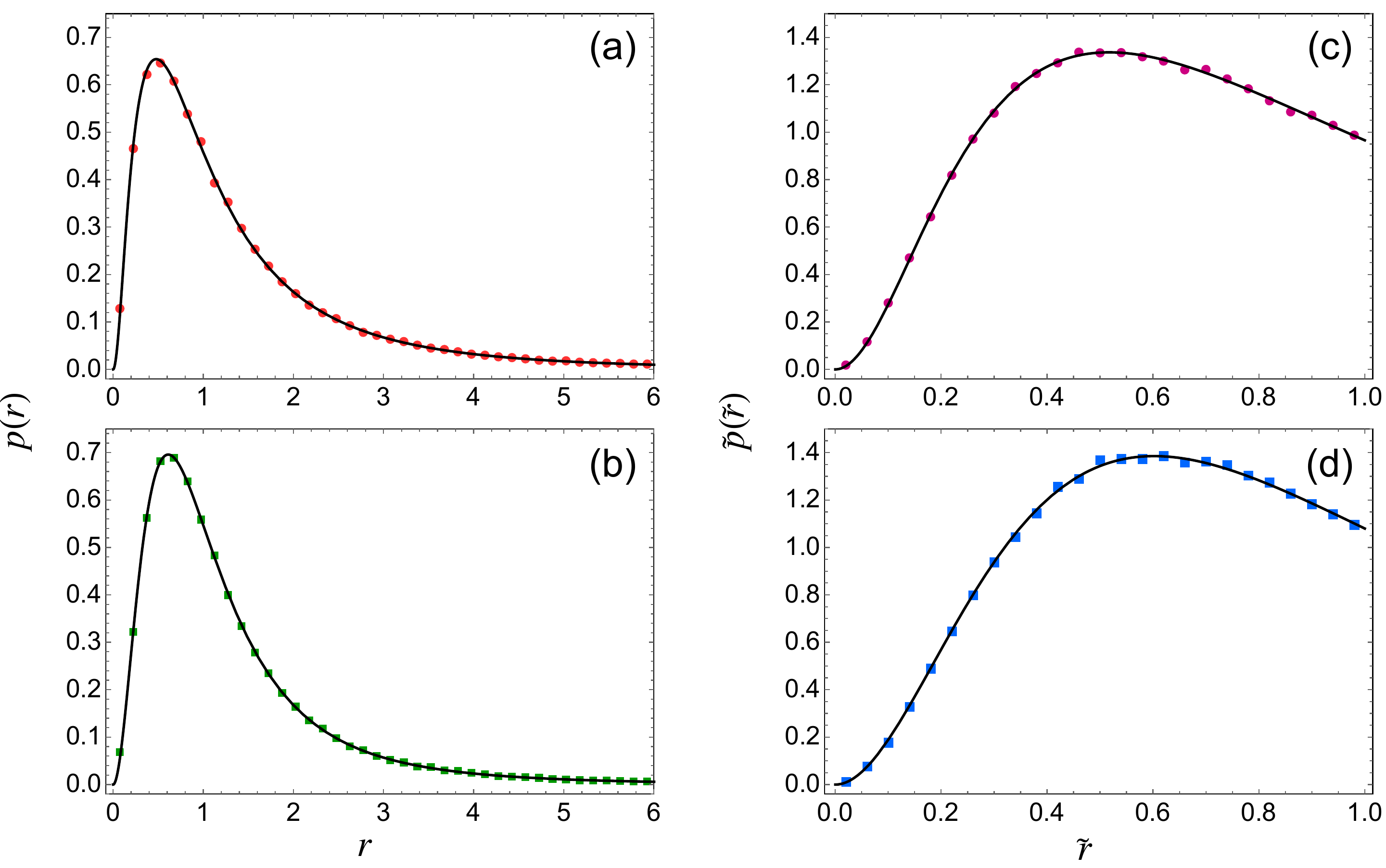}
\caption{Comparison between analytical results (solid lines) as in Eqs.~\eqref{pr1}, and~\eqref{prtilde} and numerical simulations (symbols) in the $3\times 3$ case for $p(r)$: (a) $\alpha=0.2$, (b) $\alpha=0.7$, and for $\tilde{p}(\tilde{r})$: (c) $\alpha=0.3$, (d) $\alpha=0.6$.}
\label{fig2}
\end{figure*}

The PDF $p(r)$ of the ratio $r$ can also be used to obtain the expression for the PDF of $\tilde{r}=\min(r,1/r)$. We have
\begin{equation}
\label{prtilde}
\widetilde{p}(\tilde{r})=\left[p(\tilde{r})+\frac{1}{\tilde{r}^2}p\left(\frac{1}{\tilde{r}}\right)\right]\Theta(1-\tilde{r}),
\end{equation}
where $\Theta(z)$ is the Heaviside step function.
As seen in Eq.~\eqref{jpdsp}, the joint density $\widehat{P}(x,y)$ of consecutive spacings is symmetric under the exchange for the $3\times 3$ Gaussian ensemble. This implies that the distributions of $r$ and $1/r$ are identical, i.e., $p(r)=p(1/r)/r^2$. Therefore, in this case, $\widetilde{p}(\tilde{r})=2p(\tilde{r})\Theta(1-\tilde{r})$, as also argued in Ref.~\cite{ABGR2013}. The same holds for all classical random matrix ensembles in the bulk of the spectrum in the large dimension limit and therefore it is also expected to apply for spectra of chaotic systems.

Along with the distributions of $r$ and $\tilde{r}$, the respective averages $\langle r \rangle$ and $\langle \tilde{r} \rangle$ have been found to be of importance in analyzing the crossover behavior. For instance, in Refs.~\cite{OH2007,PH2010,KRBL2010,CARK2012,CFIM2012,IORH2013,CK2013,CDK2014,LLA2015,BG2016,RCCKLR2017,GNRV2019}, the variation in these averages has been studied as certain physical parameters are varied. The averages have also been used to decide a critical value of the scaled transition parameter signifying the extent of symmetry crossover~\cite{CDK2014}. This kind of statistics is particularly useful in real complex systems where one is unaware of the full analytical solution of crossover from one symmetry class to another. In the present case, it is possible to obtain exact analytical results for the averages $\langle r \rangle$ and $\langle \tilde{r} \rangle$ using the corresponding PDFs. We have
\begin{align}
\nonumber
&\langle r \rangle = \int\limits_{0}^{\infty} r\, p(r) dr  \\ 
\nonumber
 &= \frac{9\sqrt{3}\,\alpha}{2 \pi (3+\alpha^{2})} -\frac{3}{4}+\frac{5+\alpha^{2}}{\pi (1-\alpha^{2})}\arctan\left(\frac{3-\alpha^{2}}{2\sqrt{3}\,\alpha}\right)\\ 
&  -\frac{7 + 5 \alpha^{2}}{2 \pi(1-\alpha^2)}\arctan(\frac{\alpha}{\sqrt{3}}).
\end{align}
\begin{align}
\nonumber
&\langle \tilde{r} \rangle = \int_0^\infty \tilde{r}\,\tilde{p}(\tilde{r}) d\tilde{r} = 2\int\limits_{0}^{1}  \tilde{r} \,p(\tilde{r}) d\tilde{r} \\ 
\nonumber
&=\frac{4 \left(2+\alpha ^2\right)}{\pi  \left(1-\alpha ^2\right)}\arctan\left(\frac{\sqrt{3}(1+\alpha ^2)}{2 \alpha }\right)\\
\nonumber
&-\frac{4 \sqrt{3}}{\pi  \left(1-\alpha ^2\right)^{3/2}} \arctan\left(\frac{\left(1-\alpha ^2\right)^{3/2}}{\alpha (3+\alpha ^2)}\right)\\
&-\frac{17+7 \alpha ^2}{\pi  \left(1-\alpha ^2\right)}\arctan\left(\frac{\alpha }{\sqrt{3}}\right)-\frac{1}{\pi}\arctan\left(\sqrt{3}\, \alpha \right).
\end{align}
Similar to the distributions, these quantities can also be written as functions of the equivalent symmetry-breaking parameter $\lambda$. For $\alpha\to 0^+$ and $\alpha \to 1$, the above expressions yield the average values pertaining to GOE and GUE~\cite{ABGR2013}, respectively,
\begin{align}
\nonumber
&\expval{r}_\mathrm{GOE}=\frac{7}{4}=1.75,\\
\nonumber
&\expval{r}_\mathrm{GUE}=\frac{27 \sqrt{3}}{8 \pi }-\frac{1}{2}\approx 1.3607;\\
\nonumber
&\expval{\tilde{r}}_\mathrm{GOE}=4-2\sqrt{3} \approx0.5359,\\
&\expval{\tilde{r}}_\mathrm{GUE}=\frac{2 \sqrt{3}}{\pi }-\frac{1}{2}\approx 0.6027.
\end{align}
In Fig.~\ref{fig1}, we plot the results given in Eqs.~\eqref{pr1} and~\eqref{prtilde} for probability density functions $p(r)$ and $\widetilde{p}(\tilde{r})$ using four $\alpha$ values. The $\alpha=0.01$ and $0.99$ curves are close to GOE and GUE results, respectively, while the $\alpha=0.22$ and $0.40$ curves depict intermediate situations. In Fig.~\ref{fig2}, we consider $\alpha=0.2,0.7$ for $p(r)$ and $\alpha=0.3,0.6$ for $\tilde{p}(\tilde{r})$ and compare analytical-results-based plots (solid curves) with numerical results (overlaid symbols) obtained from the Monte Carlo simulation involving 50000 matrices of dimension $3 \times 3$ following the matrix model of Eq.~(\ref{PMmatrix1}). We find excellent matches as expected.

\begin{figure*}[!ht]
\centering
\includegraphics[width=0.95\linewidth]{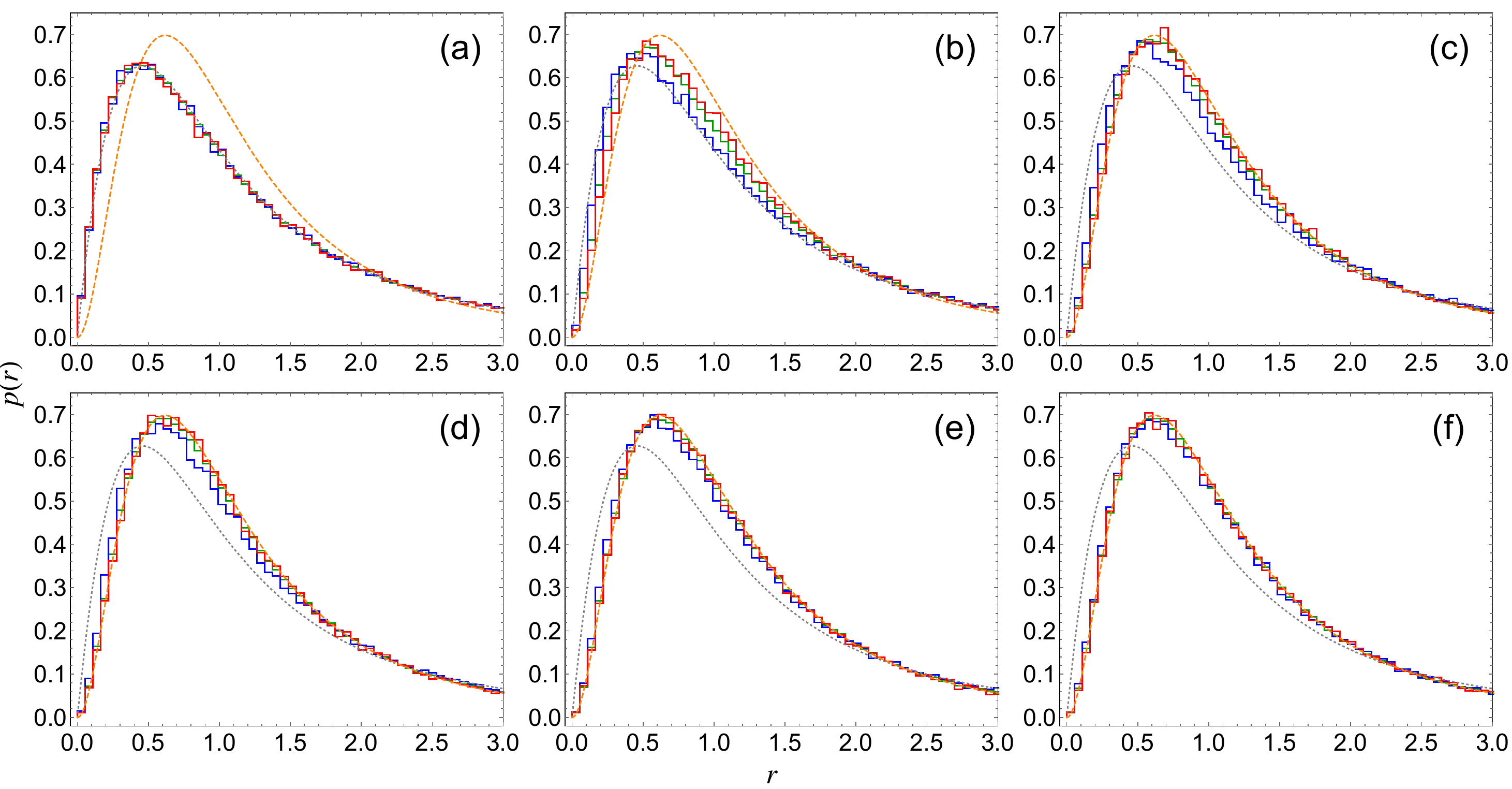}
\caption{Behavior of the ratio distribution $p(r)$ computed using the middle part (red stairs), edges (blue stairs), and the entire spectrum (green stairs) for the GOE-GUE crossover: (a) $\lambda=0$, (b) $\lambda=0.02$, (c) $\lambda=0.04$, (d) $\lambda=0.05$, (e) $\lambda=0.06$, (f) $\lambda=0.08$. The extremes corresponding to the GOE and GUE cases are depicted using dotted gray and dashed orange curves, respectively.}
\label{fig3}
\end{figure*}

\begin{figure}[!tbp]
\centering
\includegraphics[width=0.9\linewidth]{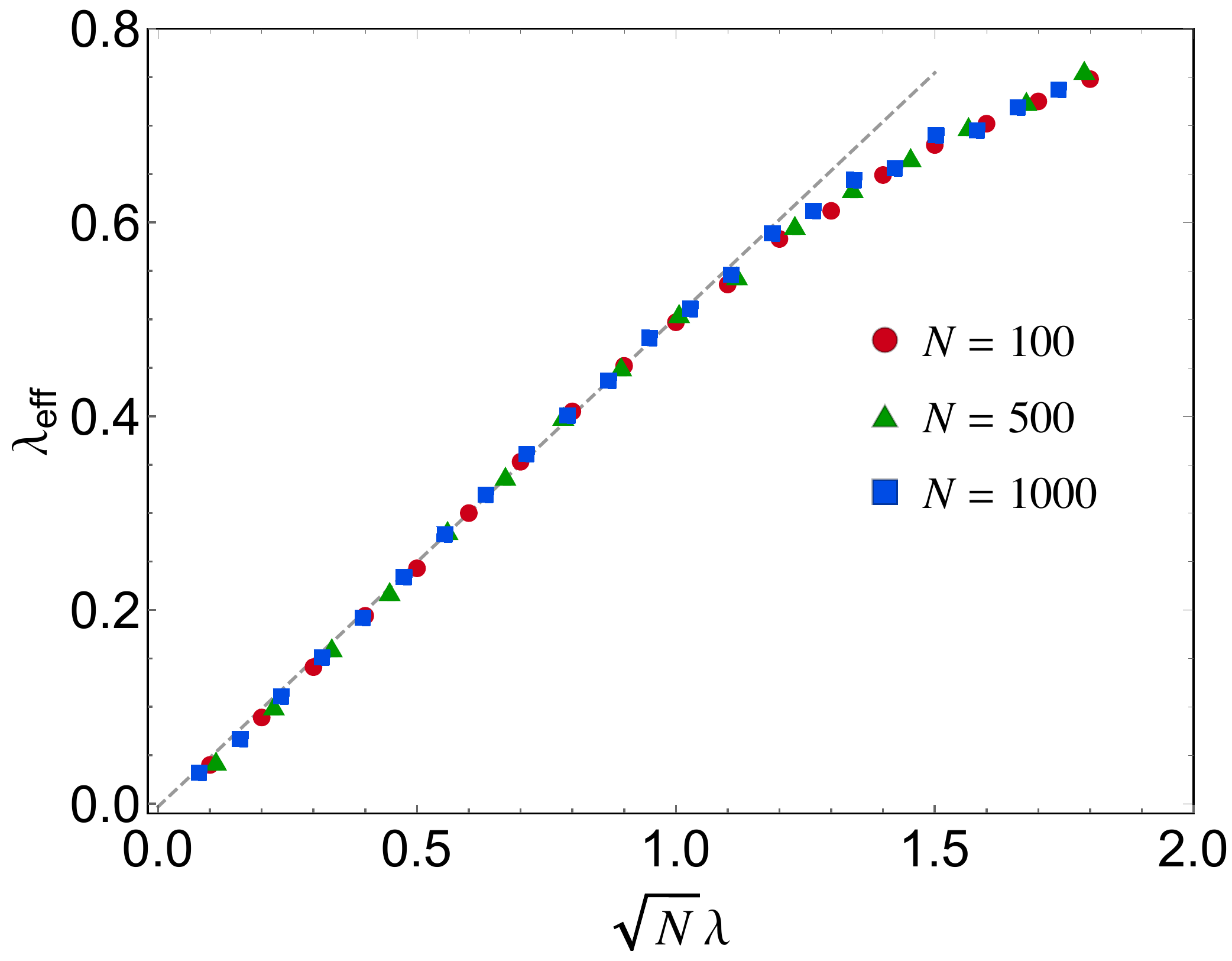}
\caption{Scaling behavior of the effective transition parameter $\lambda_\text{eff}$ in the GOE-GUE crossover. The dashed gray line is a fit based on the data points occurring in the linear regime.}
\label{FigScalingG}
\end{figure}

\section{Scaling of $\lambda$ for large $N$ Gaussian ensemble}
\label{scaling}

For the invariant cases, i.e., $\lambda=0 \,(\alpha=0)$ and $\lambda\to\infty \,(\alpha\to 1)$, it has been shown that the Wigner-surmise-like results, given by Eqs.~\eqref{pGOE} and~\eqref{pGUE}, for the ratio distribution work even for large matrix dimension $N$~\cite{ABGR2013,ABGVV2013}. However, for the intermediate cases, we must scale the crossover parameter $\alpha$ or $\lambda$ appropriately so that Eq.~\eqref{pr1} can be used for large $N$ also. As a matter of fact, it is known that the transition rate depends on the local density of states or the level density~\cite{PM1983, MP1983,Pandey1981,FKPT1988,Guhr1996,KP2009,KP2011,SBW2012}. For large $N$, the matrix model in Eq.~(1) leads to the the Wigner's semicircular level density,
\begin{equation}
R_1(x)=\begin{cases}\frac{1}{\pi}\sqrt{2N-x^2}, & -\sqrt{2N}\le x\le \sqrt{2N},\\
0, & \text{otherwise},
\end{cases}
\end{equation}
for the choice $v^2=1/[2(1+\alpha^2)]=(1+\lambda^2)/[2(1+2\lambda^2)]$. It should be noted that the mean level spacing is given by $1/R_1(x)$.
The effective crossover parameter for small $\alpha$ is $\alpha_\text{eff}\sim \alpha R_1(x)$ and equivalently, for small $\lambda$, we have $\lambda_\text{eff}\sim \lambda R_1(x)$. This implies that the transition rate is faster in the center of the semicircle. Therefore, although the ratio of consecutive spacings is independent of the local density of states, for the crossover ensemble, the fluctuation behavior of the part of spectrum near the center is closer to GUE than that near the edges. We verify this by considering 1000 matrices of size $N=1000$ using the matrix model~\eqref{PMmatrix2} and numerically obtain the ratio distribution independently using (i) 100 eigenvalues in the center of the spectrum, (ii) 100 eigenvalues comprising 50 and 50 from the left and right edges of the spectrum, and (iii) the entire spectrum. These are displayed in Fig.~\ref{fig3} using histogram stairs of colors red, blue, and green, respectively. The extremes corresponding to the GOE and GUE cases based on Eqs.~\eqref{pGOE} and~\eqref{pGUE} are shown using dotted gray and dashed orange curves, respectively. For $\lambda=0$, all histograms fall on the GOE curve, indicating that everywhere in the spectrum, the fluctuations conform to the orthogonal symmetry. Similarly, for higher values of $\lambda$, e.g., 0.08, the histograms more or less coincide with the GUE curve, implying that the crossover is almost complete. However, for intermediate cases, the red histogram (spectrum bulk) is closer to the GUE curve than the blue histogram (spectrum edges). This effect can be seen prominently in $\lambda=0.02$ and 0.04 cases. We also see that for these intermediate cases, the green histogram, which is based on the entire spectrum, is closer to the red histogram. Therefore, the overall behavior is dominated by the bulk and, as an approximation, we may use the scaling $\lambda_\text{eff}\sim \lambda R_1(0)\sim \sqrt{N} \lambda$ and consider the entire spectrum to examine the local fluctuations. To justify this proposition, based on the matrix model~\eqref{PMmatrix2}, we numerically obtain ratio distributions for matrix dimensions $N=100, 500,$ and 1000 along with several values of $\lambda$. These empirical distributions are then fitted with Eq.~\eqref{pr1} to obtain the effective crossover parameter $\lambda_\text{eff}$. The plots of $\lambda_\text{eff}$ vs $\sqrt{N} \lambda$, as shown in Fig.~\ref{FigScalingG}, do exhibit linear behavior for small $\lambda$ values and fall nearly on each other, thereby supporting our deduction above. This scaling was also found to hold in the NNSD expression for the Poisson-GOE crossover studied in Ref.~\cite{LZS1991}. Scaling behavior has also been investigated in Ref.~\cite{CDK2014} in the context of ratio distribution.

\begin{figure*}
\centering
\includegraphics[width=0.9\linewidth]{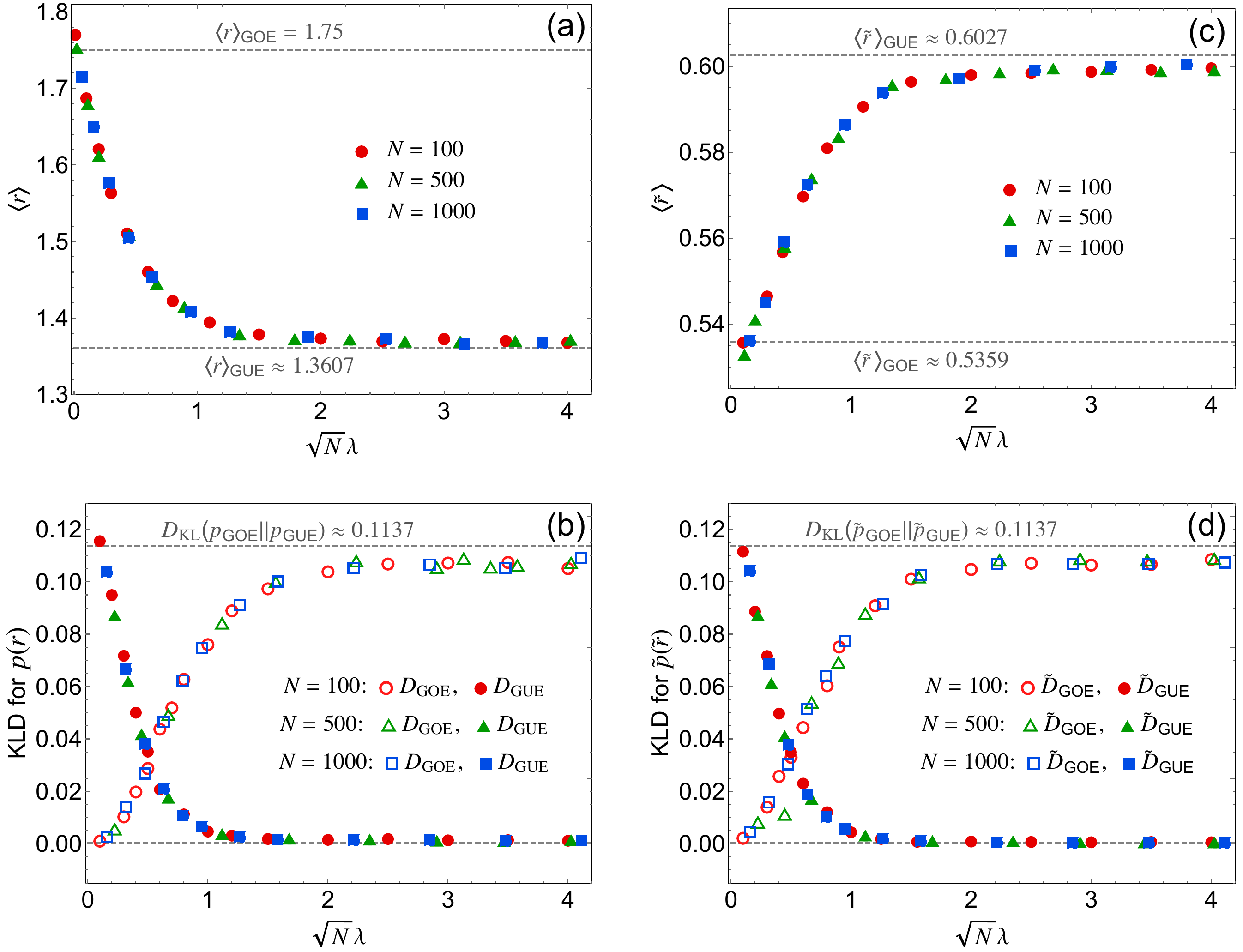}
\caption{(a),(c) Averages and (b),(d) KL divergences for $r$ and $\tilde{r}$, respectively, in GOE-GUE crossover plotted against the effective transition parameter $\sqrt{N}\lambda$. The numerical simulations have been performed using the matrix model in Eq.~\eqref{PMmatrix2} with matrix dimensions 100, 500, and 1000.} 
\label{GaussAvKLD}
\end{figure*}
\begin{figure*}[!ht]
\centering
\includegraphics[width=0.9\linewidth]{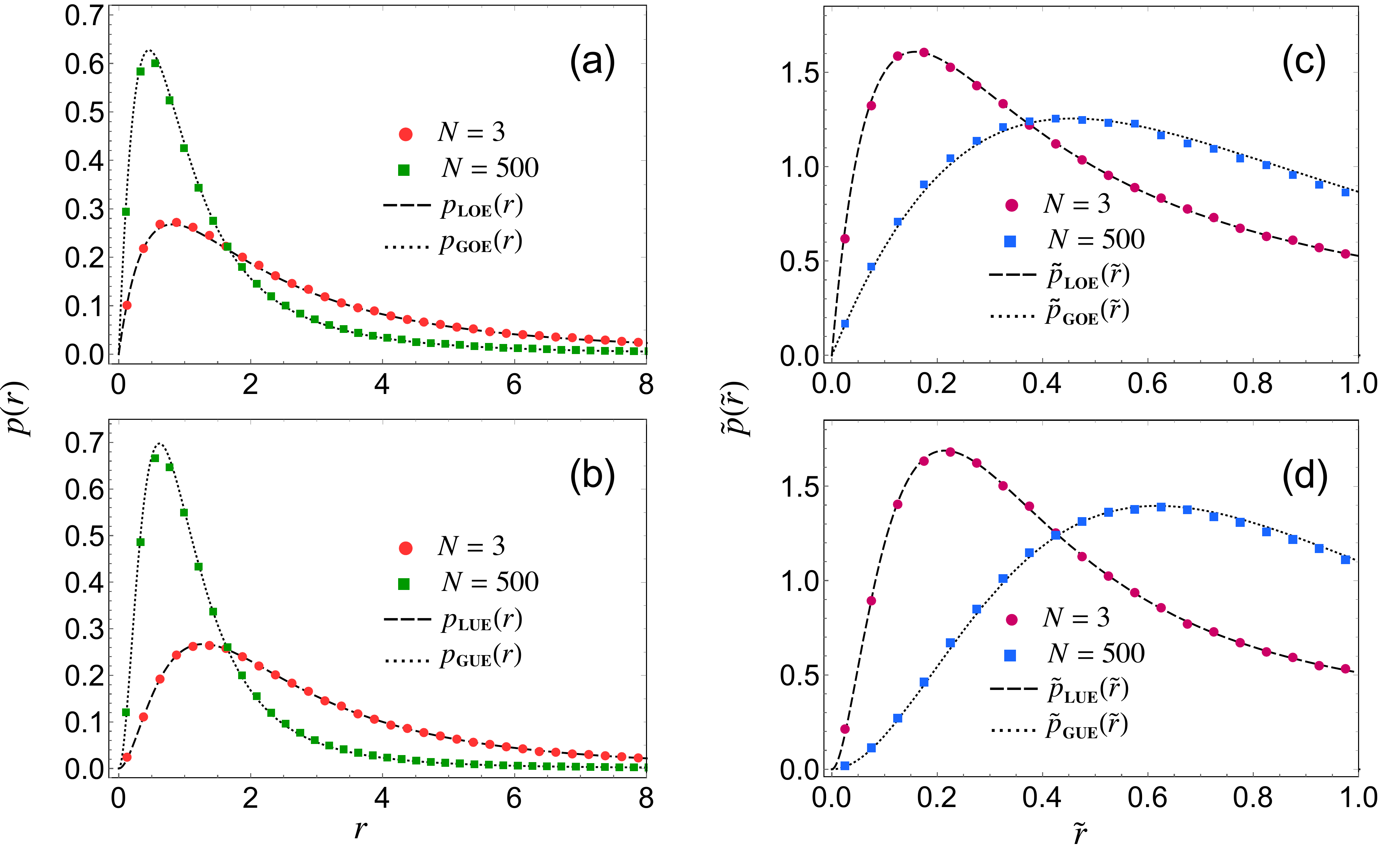}
\caption{Ratio distribution $p(r)$ for (a) Laguerre orthogonal ($\beta=1$) and (b) Laguerre unitary ($\beta=2$) ensembles. (c),(d) The respective plots for $\tilde{p}(\tilde{r})$. The dashed and dotted lines are based on $3\times 3$ analytical results for Laguerre and Gaussian ensembles, respectively. The symbols depict the simulation results for Laguerre ensembles with matrix dimensions as indicated.}
\label{FigRL}
\end{figure*}
\begin{figure}[!tbp]
\centering
\includegraphics[width=0.9\linewidth]{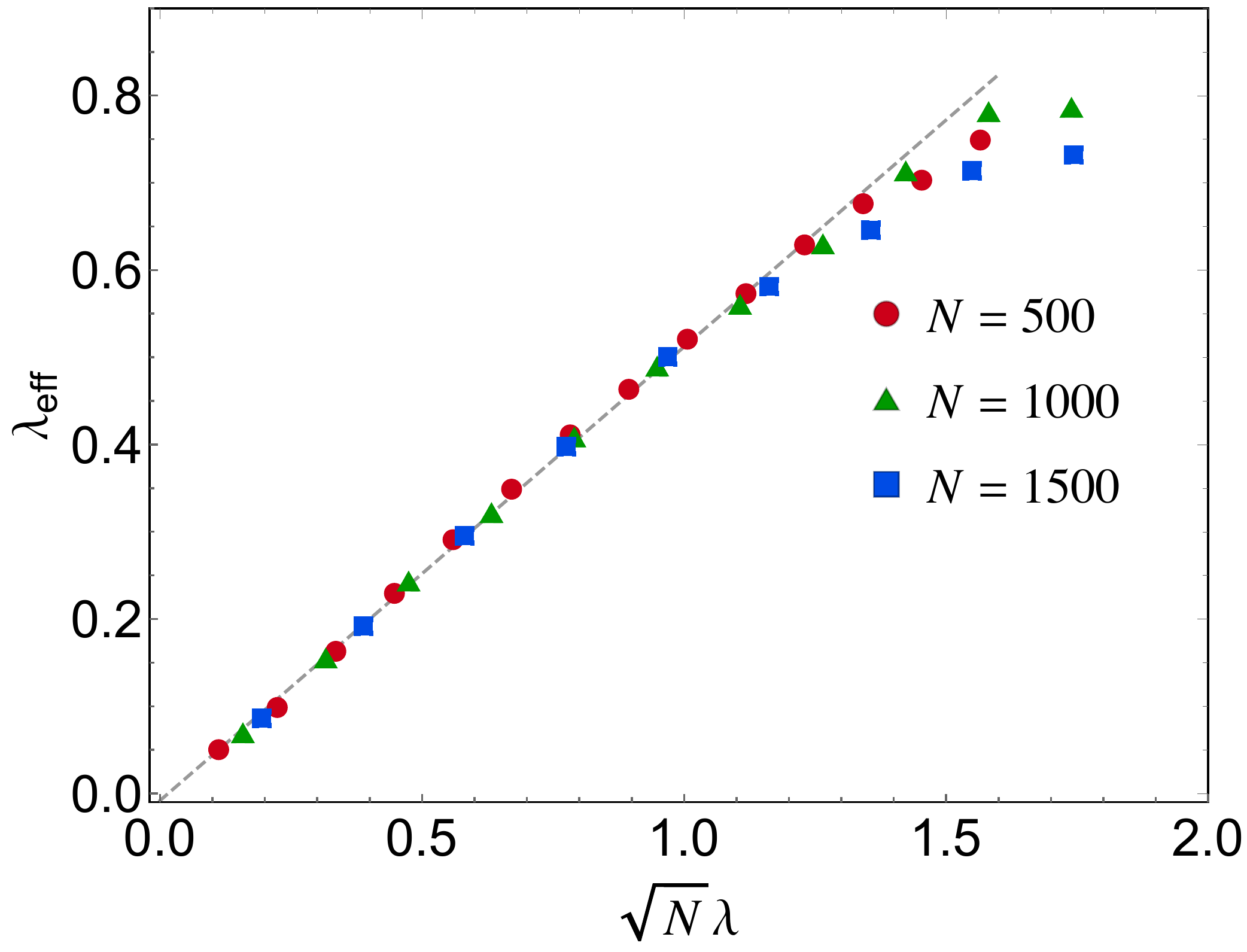}
\caption{Scaling behavior of the effective transition parameter $\lambda_\text{eff}$ in the LOE-LUE crossover. The dashed gray line is a fit based on the points occurring in the linear regime.}
\label{FigScalingL}
\end{figure}
As mentioned above, the variation of the averages $\expval{r}$ and $\expval{\tilde{r}}$ as a function of the scaled parameter has been used in earlier works~\cite{OH2007,PH2010,KRBL2010,CARK2012,CFIM2012,IORH2013,CK2013,CDK2014,LLA2015,BG2016,RCCKLR2017,GNRV2019} to visualize the crossover in spectral fluctuations. It has also been used to determine a threshold value of the scaled transition parameter above which the crossover can be deemed very nearly complete. However, since one should be careful in drawing conclusions based solely on the behavior of the average, we examine the Kullback-Leibler divergence (KLD)~\cite{KL1951} to ensure that the crossover behavior of the average is consistent with that of the full probability density function. Kullback-Leibler divergence, also called the relative entropy, serves as a measure for comparing two distributions. We use the symmetrized KLD between two PDFs, $p_1(x)$ and $p_2(x)$, which is defined as
\begin{align}
\label{KLD}
\nonumber
D_\mathrm{KL}(p_1||p_2)&=\int p_1(x)\ln\left(\frac{p_1(x)}{p_2(x)}\right)dx\\
&+\int p_2(x)\ln\left(\frac{p_2(x)}{p_1(x)}\right)dx,
\end{align}
where the integrals are over the domain on which the distributions are defined. We examine below how the empirical PDFs and the corresponding averages display crossover from GOE to GUE as the scaled transition parameter $\sqrt{N}\lambda$ is varied. We consider both $D_\mathrm{GOE}\equiv D_\mathrm{KL}(p_\mathrm{emp}||p_\mathrm{GOE})$ and $D_\mathrm{GUE}\equiv D_\mathrm{KL}(p_\mathrm{emp}||p_\mathrm{GUE})$, where $p_\mathrm{emp}$ is the empirical PDF obtained using numerical simulation. Similarly, we consider $\widetilde{D}_\mathrm{GOE}\equiv D_\mathrm{KL}(\tilde{p}_\mathrm{emp}||\tilde{p}_\mathrm{GOE})$ and $\tilde{D}_\mathrm{GUE}\equiv D_\mathrm{KL}(\tilde{p}_\mathrm{emp}||\tilde{p}_\mathrm{GUE})$. It can be calculated that $D_\mathrm{KL}(p_\mathrm{GOE}||p_\mathrm{GUE})=D_\mathrm{KL}(\tilde{p}_\mathrm{GOE}||\tilde{p}_\mathrm{GUE})\approx 0.1137$. Therefore, for the GOE-GUE crossover, $D_\mathrm{GOE}$ and $\widetilde{D}_\mathrm{GOE}$ should move away from zero to this value, while $D_\mathrm{GUE}$ and $\widetilde{D}_\mathrm{GUE}$ should approach zero starting close to this value. In practice, since the $p_\mathrm{emp}$ is obtained numerically, we use the discretized version of Eq.~\eqref{KLD}, i.e.,
\begin{align}
\label{KLD2}
\nonumber
D_\mathrm{KL}(p_1||p_2)&\approx\sum_i p_1(x_i)\ln\left(\frac{p_1(x_i)}{p_2(x_i)}\right)\Delta x\\
&+\sum_i p_2(x_i)\ln\left(\frac{p_2(x_i)}{p_1(x_i)}\right)\Delta x.
\end{align}
Moreover, for $r$, we consider the domain $[0,30]$ with bin width $\Delta r=0.06$ for extracting the numerical PDFs as the full domain $[0,\infty)$ cannot be sampled. It should be noted that for $r\in[0,30]$, theoretically we obtain $D_\mathrm{KL}(p_\mathrm{GOE}||p_\mathrm{GUE})\approx 0.1091<0.1137$. For $\tilde{r}$, we sample the full domain $[0,1]$ with $\Delta \tilde{r}=0.002$. Further, we add very small numerical values to both $p_1(x)$ and $p_2(x)$ in Eq.~\eqref{KLD2} to avoid incidences such as division by 0 or logarithm of 0. Finally, for each value of $N$, the number of matrices used for simulation was kept so as to ensure about 150000 data points for obtaining the PDFs and averages. Although the calculated values for KLD do vary based on these small numerical details, the overall behavior remains unaffected. 

In Fig.~\ref{GaussAvKLD}, we show the crossover curves for the averages and the KLDs with respect to the scaled crossover parameter for $N=100, 500,$ and $1000$. It can be seen that averages do behave in harmony with the KLDs and that the GOE-GUE crossover is almost complete for $\sqrt{N}\lambda \approx 1.5$. We also note from Fig.~\ref{FigScalingG} that up to roughly this value, there is a linear relationship between $\lambda_\text{eff}$ and $\sqrt{N}\lambda$.


\section{Application to Laguerre ensemble}
\label{ALE}

In this section, we discuss the orthogonal to unitary crossover in the Laguerre ensemble, which is also referred to as the Wishart or Wishart-Laguerre ensemble. The Laguerre ensemble appears naturally in several contexts, for example, in mathematical statistics, classical and quantum information theory~\cite{TV2004, CD2011,KP2010c,ZS2001,SZ2004,BL2002,Majumdar2011,KP2011a,KSA2017}, quantum chromodynamics~\cite{SV1993,Verbaarschot1994,GW1997,Verbaarschot2011,Akemann2017}, econophysics~\cite{LCBP1999, PGRAS1999, MS1999}, etc. We study the Laguerre ensemble below in connection with the universality of the ratio distribution obtained in the preceding section since it constitutes a very important non-Gaussian random matrix ensemble. We should also point out that very recently, in Ref.~\cite{BTS2018}, the ratio distribution was used to investigate the spectra of empirical correlation matrices for which the Laguerre ensemble serves as a natural random matrix model.

Consider ($N\times M$)-dimensional real and complex Ginibre matrices $A_1$ and $A_2$, respectively, from the distributions
\begin{equation}
P_\beta(A_\beta)\propto \exp(-\frac{\beta}{2}\tr A_\beta A_\beta^\dag);~~\beta=1,2,
\end{equation}
where $\dag$ represents conjugate transpose and may be replaced by only transpose ($T$) for $\beta=1$.
Considering $N\le M$, the Laguerre orthogonal ensemble (LOE) and Laguerre unitary ensemble (LUE) comprise the Wishart matrices $A_1A_1^T$ and $A_2A_2^\dag$, respectively. The distribution of the ratio of two consecutive level spacings is known from Ref.~\cite{ABGVV2013} for a $3\times 3$ matrix model of Laguerre ensemble in the invariant cases with Laguerre weight $e^{-\beta x/2}$. These correspond to the Wishart matrices $A_1A_1^T$ with $N=3, M=4$, and $A_2A_2^\dag$ with $N=M=3$, and read
\begin{equation}
\label{pLOE}
p_\mathrm{LOE}(r)=32\frac{\left(r^2+r\right)}{(r+2)^5},
\end{equation}
\begin{equation}
\label{pLUE}
p_\mathrm{LUE}(r)=420\frac{ \left(r^2+r\right)^2}{(r+2)^8}.
\end{equation}
 In the limits $r\to0$ and $r\to\infty$, the asymptotic behaviors of these functions are $r^{\beta}$ and $r^{-\beta-2}$, respectively, which coincide with the corresponding asymptotic behaviors of the GOE and GUE results in Eqs.~\eqref{pGOE} and \eqref{pGUE}. However, the overall shapes of these distributions based on the $3\times 3$ cases of Laguerre and Gaussian ensembles are very distinct. Furthermore, it may be shown that the joint density of consecutive spacings for the $3\times 3$ model of the Laguerre case does not have the symmetry as exhibited by the corresponding Gaussian case, as in Eq.~\eqref{jpdsp}. Therefore, the probability density $\tilde{p}(\tilde{r})$ in this case does not reduce to $2p(\tilde{r})\Theta(1-\tilde{r})$ and has to be calculated using~\eqref{prtilde}.
 
For large $N$, we find that the observed ratio distributions for the Laguerre ensemble are closely approximated by the Gaussian ensemble results instead of the above two equations. The same holds for the restricted distribution $\tilde{p}(\tilde{r})$. This is demonstrated in Fig.~\ref{FigRL} where we plot the ratio distributions obtained from numerical simulation based on the above-described Wishart-Laguerre matrix model. For comparison, we also plot the analytical expressions from Eqs.~\eqref{pGOE},~\eqref{pGUE},~\eqref{pLOE}, and~\eqref{pLUE}. We can see that for $N=500$, the empirical ratio distributions for the Laguerre ensemble are well described by the Gaussian $3\times3$ results, thereby confirming their universality. This is similar to the Wigner surmise based on $2\times2$ matrices from the Gaussian ensemble. The underlying reason is that in the bulk of the spectrum of a large class of random matrix ensembles, the correlation functions are governed by the sine kernel~\cite{Mehta2004,Forrester2010,DKMVZ1999,PG2001,GP2002,DGKV2007}. The exact distribution of the nearest neighbor can then be expressed in terms of Fredholm eigenvalues of this kernel, which in turn is very well approximated by the Wigner surmise~\cite{Mehta2004}. A similar explanation holds behind the universality of the ratio distribution, as argued in Ref.~\cite{ABGR2013}.

We now analyze the LOE-LUE crossover. The matrix model that is implemented is
\begin{equation}
\label{wishart}
W=AA^\dag,
\end{equation}
where 
\begin{equation}
A=\frac{1}{\sqrt{1+\lambda^2}}A_1+\frac{\lambda}{\sqrt{1+\lambda^2}}A_2.
\end{equation}
The limits $\lambda\to0$ and $\lambda\to\infty$ produce LOE and LUE, respectively.
We focus on the large $N=M$ case, for which the level density for the Wishart random matrix $W$ is given by
\begin{equation}
R_1(x)=\begin{cases}\frac{1}{\pi}\sqrt{\frac{2N-x}{x}}, & 0\le x\le 2N,\\
0, & \text{otherwise}.
\end{cases}
\end{equation}
For the above LOE to LUE crossover ensemble, the effective transition parameter is known to be $\lambda_\text{eff}\sim\lambda\sqrt{x}R_1(x)\sim \lambda\sqrt{2N-x}$~\cite{KP2009,KP2011,footnote}, which depends on the location in the spectrum. However, away from the right edge, we have $\lambda_\text{eff}\sim\sqrt{N}\lambda$ and we expect this scaling to work for the entire spectrum collectively, similar to the Gaussian case. We verify this with the aid of matrices generated using Eq.~\eqref{wishart} for $N=500,1000,1500$ and varying $\lambda$ values. The empirical ratio distributions are fitted with the formula in Eq.~\eqref{pr1} to obtain the effective $\lambda_\text{eff}$ and then plotted against $\sqrt{N}\lambda$ in Fig.~\ref{FigScalingL}. The data points almost fall on a common line for small values of $\lambda$, which indicates the validity of the above-mentioned scaling for LOE-LUE crossover.

\begin{figure*}
\centering
\includegraphics[width=0.9\linewidth]{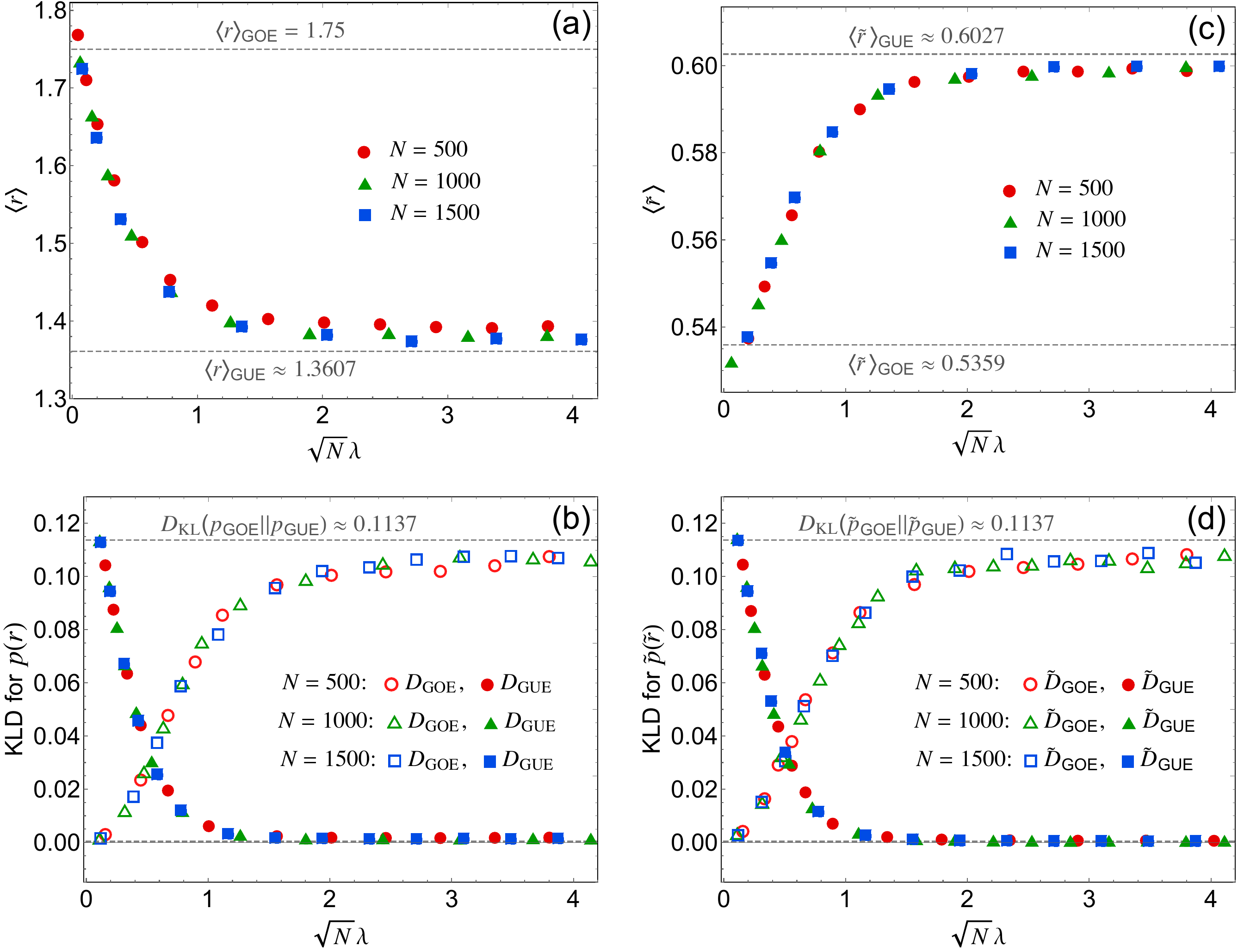}
\caption{(a),(c) Averages and (b),(d) KL divergences for $r$ and $\tilde{r}$, respectively, in LOE-LUE crossover plotted against the effective transition parameter $\sqrt{N}\lambda$. The numerical simulations are based on the matrix model in Eq.~\eqref{wishart} for matrix dimensions 500, 1000, and 1500.}
\label{LagAvKLD}
\end{figure*}

We also investigate the behavior of the averages and KL divergences in the present case similar to the Gaussian ensemble. These are shown in Fig.~\ref{LagAvKLD} and we find that by the value $\sqrt{N}\lambda\approx 1.5$, the crossover is nearly achieved. It should be noted that $\lambda_\text{eff}$ and $\sqrt{N}\lambda$ behave linearly with respect to each other up to this value, as observed in Fig.~\ref{FigScalingL}.


\section{Symmetry crossover in quantum kicked rotor}
\label{QKR}
\begin{figure}[!tbp]
\centering
\includegraphics[width=0.9\linewidth]{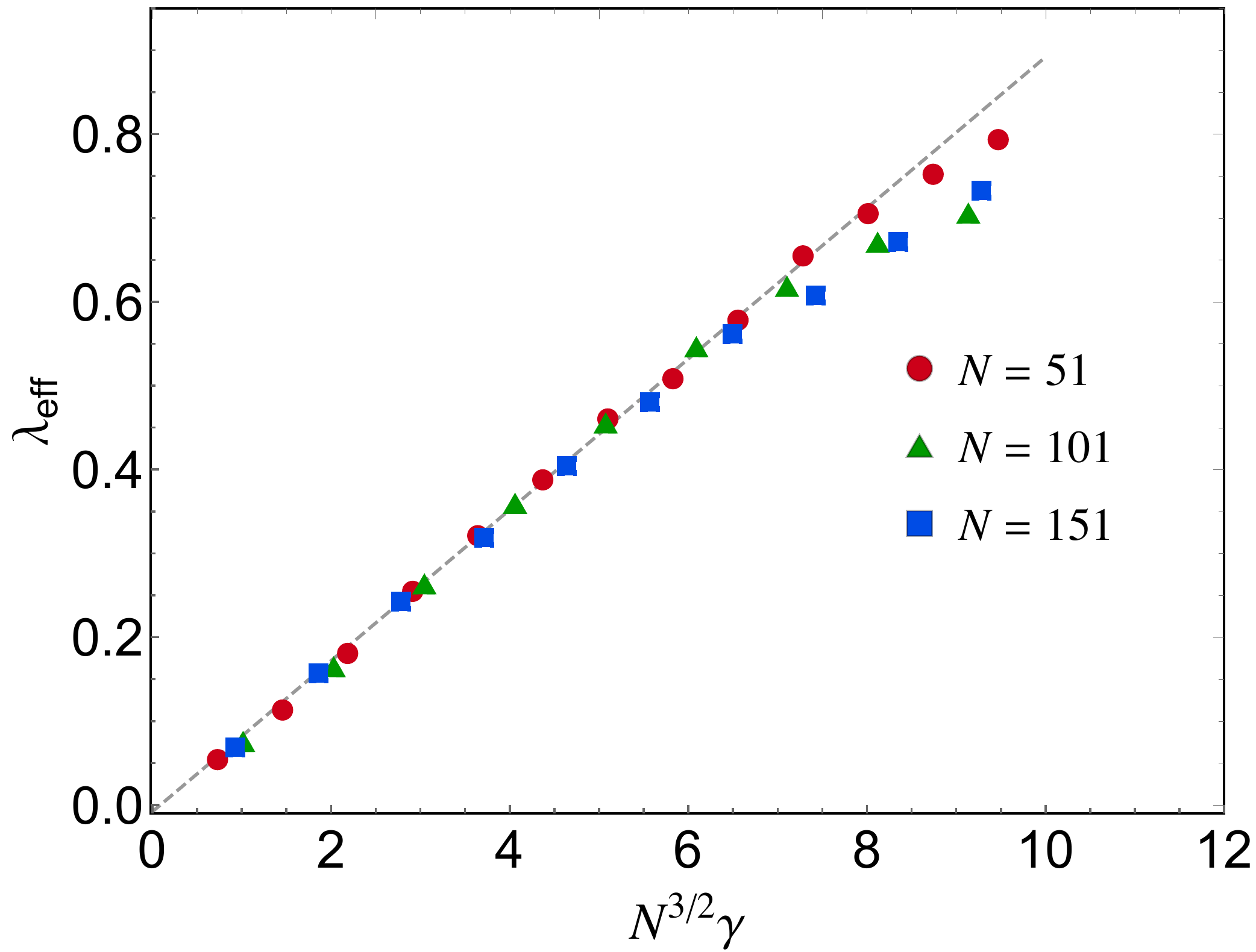}
\caption{Dependence of the effective crossover parameter $\lambda_\text{eff}$ on the time-reversal violation parameter $\gamma$ of QKR system.The dashed gray line is a fit based on the points occurring in the linear regime.}
\label{FigscalingQKR}
\end{figure}
\begin{figure*}[!ht]
\centering
\subfloat{\includegraphics[width=0.9\linewidth]{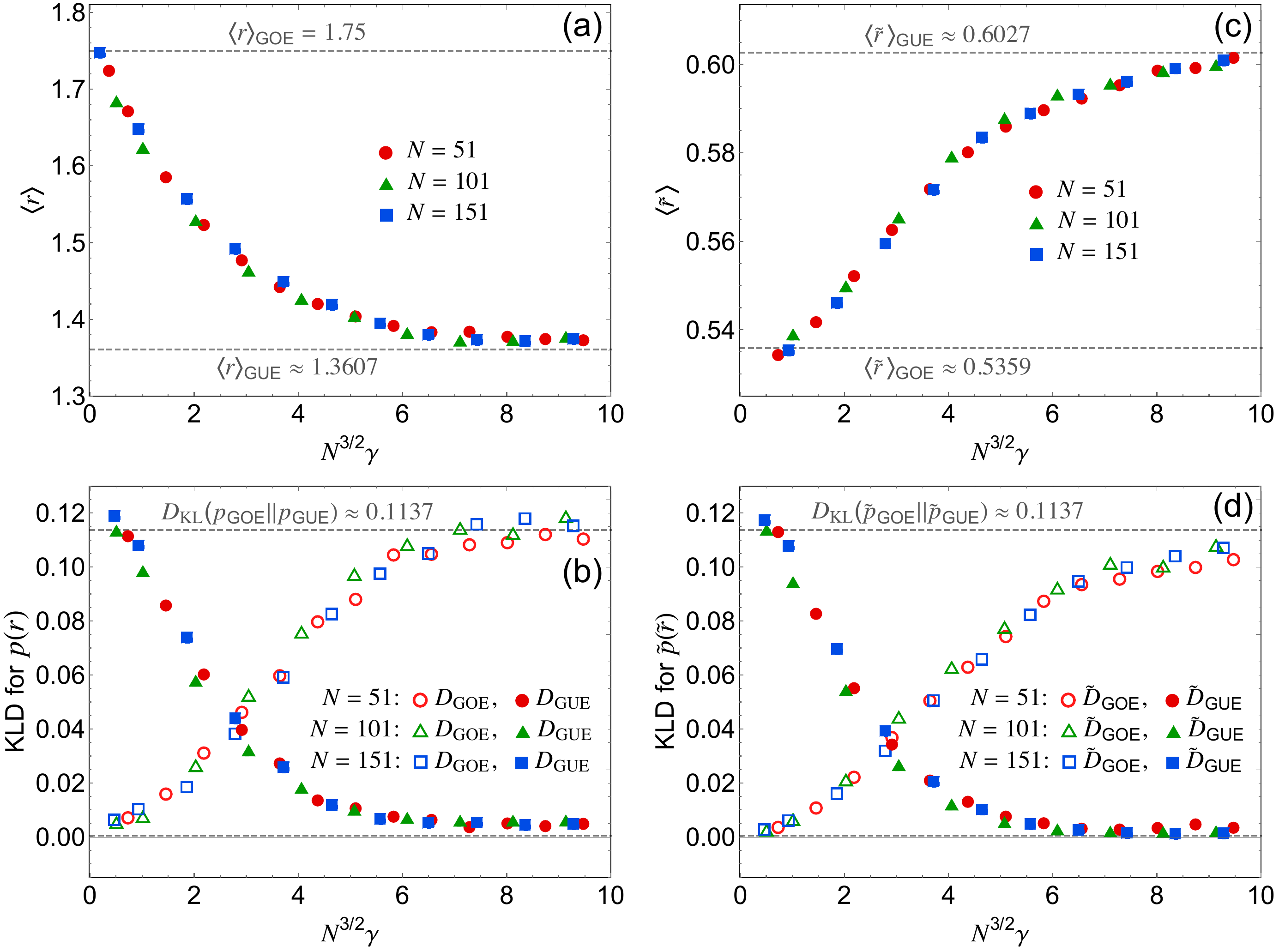}}
\caption{(a),(c) Averages and (b),(d) KL divergences for $r$ and $\tilde{r}$, respectively, in the quantum kicked rotor system plotted against the effective transition parameter $N^{3/2}\gamma$ for Floquet operator dimensions 51, 101, and 151.}
\label{KRAvKLD}
\end{figure*}

The quantum kicked rotor (QKR) was introduced to serve as a simple yet significant model to aid investigations into quantum chaos~\cite{CCFI1979}. Since then, QKR and its variants have been extensively used in several contexts~\cite{CCFI1979,Izrailev1986,Casati1990,Izrailev1990,PRS1993,SP1997,AL2001,SL2015,PKP2017,KPP2019,JPP2019}. The Hamiltonian for the kicked rotor system is given by
\begin{equation}
\mathcal{H}=\frac{(p+\gamma)^2}{2}+\alpha \cos(\theta+\theta_0)\sum_{n=-\infty}^\infty \delta(t-n).
\end{equation}
It describes a particle restricted to a ring and receiving position-dependent periodic kicks. In the above equation, $p$ is the momentum operator, $\theta$ is the position operator, and $\alpha$ is the stochasticity parameter. The role of the parameters $\gamma$ is to imitate the effect of time-reversal breaking, while $\theta_0$ facilitates the parity violation. The periodic kicks at integer time instants ($t=n$) are provided by the Dirac comb.
The associated quantum dynamics may be described by the discrete time evolution operator (Floquet operator) $U$ by imposing the torus boundary condition on the phase space. Considering the finite-dimensional model, we obtain the evolution operator in position basis as~\cite{Casati1990,Izrailev1986}
\begin{align}
U_{mn} &= \frac{1}{N} \exp[-i \alpha \cos(\frac{2 \pi m}{N}+\theta_{0})]\\ \nonumber& \times \sum_{l=-N'}^{N'}\exp [-i\left(\frac{l^2}{2}-\gamma l+ \frac{2\pi l(m-n)}{N}\right)].
\end{align} 
In the above summation, $N'$ is $ (N-1)/2 $ with $N$ odd and $m,n$ take the values $-N',-N'+1,...,N'$. A high degree of chaos is necessary for the spectral fluctuation properties of the QKR to correspond to classical RMT ensembles~\cite{Casati1990, Izrailev1986, Izrailev1990}. This can be achieved by assigning the parameter $\alpha$ a very high value. For studying the violation of time-reversal invariance, we set $\theta_0=\pi/(2N)$ (fully broken parity symmetry) and then vary $\gamma$ gradually. This leads the eigenangle (or eigenphase) spectrum of $U$ to exhibit a transition from orthogonal to unitary class, which has been shown to be described very well using the crossover from circular orthogonal ensemble (COE) to circular unitary ensemble (CUE)~\cite{PS1991,PRS1993,SP1997}. In the following, we examine the crossover in the distribution of the ratio of two consecutive eigenangles and dependence of the transition parameter $\lambda_\text{eff}$ on the time-reversal violation parameter $\gamma$.

For our analysis, we consider $N = 51, 101,151$ and generate the corresponding ensembles of $U$ matrices by varying the stochasticity parameter $\alpha$ in the neighborhood of 20000. These matrices are diagonalized to obtain the eigenvalues which are of the form of $e^{i\phi_j}$, with $\phi_j$ being the eigenangles. The eigenangles are uniformly distributed in $[-\pi,\pi)$ and therefore the level density is $R_1(\phi)=N/(2\pi)$. It is also known based on a semiclassical analysis that the effective transition parameter $\lambda_\text{eff}$ for this system is proportional to $\gamma N^{3/2}$ \cite{PRS1993, SP1997}. This is verified in Fig.~\ref{FigscalingQKR} where we plot $\lambda_\text{eff}$ versus $N^{3/2}\gamma$. The $\lambda_\text{eff}$ values have been obtained by fitting the numerically obtained ratio distributions with Eq.~\eqref{pr1}. We observe linear behavior for small values of $\gamma$, which confirms the above relationship between $\lambda_\text{eff}$ and $\gamma$. 

Finally, we examine the transition behavior of the averages and KL divergences with the scaled crossover parameter. The corresponding results are shown in Fig.~\ref{KRAvKLD} and we find that the crossover is nearly complete for $N^{3/2}\gamma\approx 7$. Moreover, as already seen in Fig.~\ref{FigscalingQKR}, up to this value the scaled crossover parameter $N^{3/2}$ varies linearly with $\lambda_\text{eff}$.

\section{Summary and Outlook}
\label{discsum}

For analyzing the spectral fluctuations of complex physical systems, distributions of nearest-neighbor spacing and their ratio are two widely used measures. The latter is convenient to apply since it circumvents the procedure of unfolding the spectrum. For the invariant classes, the ratio distributions were recently derived. These results are analogous to the Wigner surmise for the nearest-neighbor spacing distributions. For the crossover ensembles, such as one describing a gradual time-reversal breaking, similar results are known for the nearest-neighbor spacing distribution. However, such results corresponding to the ratio of spacings have been missing. In this work, we aimed to fill this gap by deriving the Wigner-surmise-like result for ratio distribution and the corresponding average in the orthogonal to unitary symmetry crossover. These results were verified using Monte Carlo simulations. We also examined the proper scaling of the transition parameter which is needed to handle large size spectra by studying the Gaussian and Laguerre crossover ensembles. Additionally, we investigated the effect of time-reversal symmetry breaking in the spectrum of quantum kicked rotor by examining the behavior of the ratio distribution of eigenangles and the associated scaling of the transition parameter.

Aside from the orthogonal-unitary crossover ensemble, there are several random matrix models which are useful in exploring other kinds of spectral transitions. Although phenomenological formulas for the ratio distribution are available for some of these, it would be of interest to see if exact analytical results can be obtained.

\acknowledgements
A.S. acknowledges DST-INSPIRE for providing the research fellowship (Grant No. IF170612). S.K.~acknowledges the support by Grant No. EMR/2016/000823 provided by SERB, DST, Government of India. The authors are also grateful to the anonymous reviewers for their constructive remarks.


\end{document}